\documentclass[twocolumn,showpacs,preprintnumbers,amsmath,amssymb]{revtex4}
\usepackage{graphics}
\usepackage{epsfig}
\usepackage{color}
\usepackage{subfigure}

\begin{document}

\title{Efficiency of cyclic devices working with non-Boltzmannian fluids: challenging the second principle of thermodynamics}
\author{D. Fanelli$^{1}$, G. De Ninno$^{2,3}$, A. Turchi$^{4}$}
\affiliation{1. Dipartimento di Energetica "S. Stecco" and INFN, University of Florence, Via S. Marta 3, 50139 Florence, Italy\\
2. Sincrotrone Trieste, S.S. 14 km 163.5, Basovizza (Ts), Italy \\
3. Physics Department, Nova Gorica University, Nova Gorica, Slovenia \\
4. Dipartimento di Sistemi e Informatica and INFN, University of Florence, Via S. Marta 3, 50139 Florence, Italy 
}

%\date{\today}

\begin{abstract}

According to classical Boltzmannian thermodynamics, the efficiency of a cyclic machine is strictly lower than one. Such a result is a straightforward  consequence of the second principle of thermodynamics. Recent advances in the study of the thermodynamics of long-range interacting system report however on a rather intricate zoology of peculiar behaviors, which are occasionally in contrast with customarily accepted scenarios, dueling with intuition and common sense. In this paper, a thermodynamical cycle is assembled 
for an ideal device working with non-Boltzmaniann long-range fluid and operating in contact with two thermal reservoirs. The system is analytically shown to violate the second principle of thermodynamics, a phenomenon that ultimately relates to the existence of regions with negative kinetic specific heat, in the canonical ensemble for the system under scrutiny. We argue that the validity of the second principle of thermodynamics can be possibly restored, by revisiting the definition of canonical ensemble, as well as the Fourier law of heat transport, and consequently relaxing the constraint on the maximal efficiency as imposed by the Carnot theorem.    
\end{abstract}

\maketitle

The second principle of thermodynamics represents a fundamental milestone of physics \cite{Zemansky}. It is a principle, and as such cannot be proven on a rigorous mathematical basis. Its validity  is however supported by an impressively large gallery of experimental facts and consequently assurged to the status of inviolable universal law. It admits different formulations, the first of which is credited to the German physicist Rudolf Clausius \cite{Clausius}: {\it No process is possible whose sole result is the transfer of heat from a body of lower temperature to a body of higher temperature.} This in turn traces back to the concept of irreversibility: the heat cannot flow from cold to hot regions without external work being performed on the system, as e.g. it happens in a refrigerator. 

An alternative formulation of the second principle due to the Irish physicist Lord Kelvin \cite{Kelvin} recites instead: {\it No process is possible in which the sole result is the absorption of heat from a reservoir and its complete conversion into work.} In practice, 
one cannot entirely transform the heat absorbed by a reservoir into work, without dissipating part of it. Rephrasing the latter statement: it is thermodynamically impossible to obtain a 100 $\%$ efficiency from a cyclic heat engine.     

Indeed, Carnot's theorem \cite{Carnot} limits the maximum efficiency $\eta$ for any possible engine, and can be seen as a direct byproduct of the second principle. All irreversible heat engines operating between two heat reservoirs, respectively characterized by thermodynamic temperatures $T_C$ (cold) and $T_H$ (hot), are less efficient than a Carnot engine working in contact with the very same reservoirs. The Carnot maximal efficiency can be straightforwardly quantified in $\eta=1-T_C/T_H$ as discussed in any basic textbooks. 

The validity of the second principle of thermodynamic reflects, as a corollary, the impossibility of realizing 
perpetual motion of the second kind. This undisputed verdict is recognized by the whole scientific community, despite 
the fact that the pursuit of perpetual motion remains a popular subject of speculations. If such a device existed, 
it could transform heat into useful work, without requiring any dissipative losses, a procedure that is ultimately against the second law. 

As already emphasized, the above picture is at core of our current understanding of physics, it appears to be solidly grounded on experiments and no doubt is cast on its correctness. Recently, however, a large body of evidence has emerged which testifies on the peculiar and rather unintuitive thermodynamic properties of long range interacting systems \cite{Loris}. Microcanonical and canonical statistical ensemble can be inequivalent, an intriguing phenomenon that has been associated to e.g. negative (microcanonical) specific heat \cite{Barre, Thirring} and susceptibility (for magnetic-like systems) \cite{Touchette, DeNinno}. In classical thermodynamics, the specific heat measures the energy required to increase the temperature of a unit quantity of a substance by a unit of (thermodynamic) temperature. As common sense dictates, providing energy to the system should induce heating, meaning an increase in the system temperature. At odd with intuition, non-Boltzmannian systems (as those displaying long range traits) can cool down when forced to increase their internal energy. The canonical specific heat is however always positive \cite{Gross}, a mathematical constraint that descends from statistical mechanics. For non-Boltzmannian systems, however, the kinetic temperature {\it differs} from the thermodynamic one, the former providing a direct measure of microscopic agitation. By elaborating on this scenario, it was demonstrated \cite{Staniscia} that negative kinetic specific heat can be {\it measured} in the canonical statistical ensemble. In fact, due to a complex self-organized interplay between collective and microscopic degrees of freedom, the particles belonging to the system under scrutiny can gain in kinetic energy when an energy quota is passed from the system to the heat reservoir. The opposite holds when the energy is instead released from the thermostat to the system. In reference \cite{Staniscia}, it was speculated that this peculiar prerogative traces back to the non-Boltzmanniann (equilibrium) character of the particles' statistics. This could eventually open up the perspective of realizing novel, more efficient
thermal devices. 

This paper is devoted to shed light onto such an issue, elucidating on a viable direction of investigation of paramount importance for both applied and fundamental research. Working within a paradigmatic mean field model, and assuming the conventional definition of the canonical ensemble, we rigorously show that a thermodynamic device operating with a non-Boltzmannian fluid can be conceived, which violates the second law of thermodynamics. The only possibility we see to reconcile our results with the second principle is to modify the Fourier law of heat conduction, by assuming that heat is transported along the gradient of the {\it kinetic temperature} and not the thermodynamic temperature, as classically believed. This modification is not innocent and can impact the upper boundary of 
thermal efficiency, as specified by Carnot theorem. In both ways, we hence argue that working with unconventional non-Boltzmannian fluids can alter our current perception of the concept of efficiency, beyond the standard Boltzmann picture to which predictive tools are at present indirectly anchored.

{\bf The long-range non Boltzmannian fluid.}

Let us now turn to introducing the general background that defines the playground arena for our thermodynamics derivation. We will deal with a simple 
one-dimensional model and label with $\theta$ the configuration variable, $p$ being its conjugate momentum. In the continuum limit, we assume the single particle distribution function $f(\theta,p,t)$ to obey to the following Vlasov equation: 
\begin{equation}
\frac{\partial f}{\partial t} + p\frac{\partial f}{\partial \theta} -
\frac{d V}{d \theta} \frac{\partial f}{\partial p}=0\quad ,
\label{eq:Vlasov}
\end{equation}
where the potential $V(\theta)[f]$ is self-consistently defined in terms of the distribution itself. The problem is hence  cast in a general setting which embraces diverse fields of application, from plasma physics to cosmology. Our declared aim is to construct and study an (equilibrium) cyclic machine working with a Vlasov fluid. To this end, we shall specialize further by assuming a specific form of the potential, which will enable us to carry out the sought calculations in details. The implications of our analysis can be however extended to other realms, provided the prototypical equation (\ref{eq:Vlasov}) holds true. In fact, equation (\ref{eq:Vlasov}) yields a Fermi like entropy functional, which embeds dynamical constraints and that can be derived from first principles through the usual counting of accessible microscopic states.  As a test case, we consider a paradigmatic example of long range systems, the so called Hamiltonian Mean Field (HMF) model, subject to an external magnetic forcing of strength $h$. The HMF is a popular model of spins, interacting all--to--all via a cosinus like potential. $V(\theta)[f]$ takes in this case the form:
\begin{equation}
V(\theta)[f] = 1 - M[f] \cos(\theta) + h \cos(\theta)  ~
\label{eq:pot_magn}
\end{equation}
where $M[f]$ stands for the magnetization: 
\begin{equation}
M[f] = \int_{-\pi}^{\pi} \int_{-\infty}^{\infty}  f(\theta,p,t) \, \cos{\theta}  {\mathrm d}\theta
{\mathrm d}p\quad , 
\label{eq:mag}
\end{equation}
which measures the degree of spatial spins' bunching. Including the kinetic contribution, the specific energy 
reads $e[f]=\int \int (p^2/{2}) f(\theta,p,t) {\mathrm d}\theta
{\mathrm d}p - ({M^2 - 1})/{2} + h M$. In its original formulation, the HMF is a discrete model composed of 
$N$ particles (spins) in mutual interaction, and has been thoroughly studied in the unperturbed limit $h=0$.
Starting from {\it out--of--equilibrium} initial conditions, the system gets trapped in 
Quasi-Stationary States (QSS) \cite{Loris}, whose lifetime diverges when increasing the number of particles
$N$. When performing the thermodynamic limit ($N \rightarrow \infty$) {\it before} the infinite time limit, the system
remains permanently confined in the QSS, which are characterized by non Gaussian distributions and hence
substantially differ from the deputed Boltzmann-Gibbs equilibrium. It has been shown that QSS's can be related to the stable (equilibrium) steady states of the Vlasov equation (\ref{eq:Vlasov}) describing the system in the continuum limit \cite{Loris} \footnote{It is worth stressing, that according to the Kac scaling prescription, the Vlasov limit is obtained by reducing the interaction strength (by a factor $1/2N$), while increasing the number of particles. As a result, the energy scales like $N$, 
and the system is extensive (but not additive).} Studying the equilibria of the Vlasov equations could be hence equivalently 
interpreted as addressing the out of equilibrium QSS dynamics of the associated HMF model. 

The microcanonical (constant energy) behavior of system (\ref{eq:Vlasov}) can be investigated via a direct statistical 
mechanics treatment, inspired to the seminal work of Lynden-Bell \cite{LyndenBell}, and originally developed in the context of stellar dynamics (see, e.g., \cite{Chavanis}). Starting from an arbitrary initial condition, the Vlasov evolution makes the single-particle distribution function stir and distort in phase space. The hypervolumes of its values $\eta$ (levels of phase density) are however conserved. Label with $\rho(\theta,p,\eta)$ the probability density of finding the level of phase density $\eta$ in a small neighborhood of the position $(\theta, p)$ in phase space. One can thus define a ``coarse-grained'' distribution function $\bar{f}(\theta,p)=\int \rho(\theta,p,\eta) \eta d \eta$. The microcanonical equilibrium distribution maximases the mixing entropy
$s=-\int \rho(\theta,p,\eta) \ln \rho(\theta,p,\eta) d \eta d \theta d p$, while conserving the invariant of the motion, as prescribed by standard statistical mechanics. For the sake of simplicity, and to allow for relatively straightforward analytical progress, we solely consider the case of a two-level water-bag initial conditions, i.e. $\eta=f_0$ or  $\eta=0$ \footnote{The adequacy of Lynden-Bell approach is corroborated through direct comparison with simulations. It can be also shown that the theory works well when generalizing the initial condition to account 
for more levels. In the following we will assume an initial water-bag distribution with zero momentum.}. Within this approximation, the entropy reads:
\begin{equation}
s[\bar{f}]=-\int \!\!{\mathrm d}p{\mathrm d}\theta \,
\left[\frac{\bar{f}}{f_0} \ln \frac{\bar{f}}{f_0}
+\left(1-\frac{\bar{f}}{f_0}\right)\ln
\left(1-\frac{\bar{f}}{f_0}\right)\right],
\label{eq:entropieVlasov}
\end{equation}
where $\bar{f}=f_0 \rho$. By maximizing the entropy functional, one  obtains the microcanonical solution in terms of
a {\it fermionic} distribution (see below),  which coincides with a particular class of stable stationary solutions of the Vlasov equation. 

Given the microcanonical entropy (that self-consistently depends on the energy $e$),  one can always calculate the canonical rescaled free energy, $\phi(\beta, \bar{f})$,  as the Legendre-Fenchel Transform (LFT): $\phi(\beta,\bar{f})=\mbox{inf}_{e}\left[\beta e -s(e,\bar{f})\right]$. This is a universally adopted procedure: $\beta$ is a free parameter equal to the inverse of the (constant, in the canonical ensemble) thermodynamic temperature of the system. Requiring the free energy to be stationary, under the dynamical constraints, one recovers a closed mathematical prediction for the particles distribution, which again is fermionic. Concretely,
the canonical distribution is:   

\begin{multline}
  \label{eq:barf}
  \bar{f}(\theta,p)= \\
  \frac{f_0}{ 1+e^{\displaystyle\beta f_0 (p^2/2 -M[\bar{f}]\cos\theta-h \cos\theta)+\alpha}}.
\end{multline} 

where $\alpha$ plays the role of a Lagrange multiplier associated to mass conservation. The latter equation is 
formally identical to the one obtained by requiring the entropy (\ref{eq:entropieVlasov}) to be stationary in the framework of the microcanonical ensemble \cite{Antoniazzi}. In that case, $\beta$ and $\alpha$ are Lagrange multipliers associated to energy and mass conservation. In both cases, the respective Lagrange multipliers can be calculated by solving the self-consistent system of equations obtained by imposing the constraints conditions. Consequently, the coarse grained distribution function $\bar{f}$ and the magnetization $M$ can be calculated, for any fixed initial condition, i.e. given $f_0$, $h$ and $e$ (resp. $\beta$) in the microcanonical (resp. canonical) setting.  
 
A remark is mandatory at this point, which will prove crucial in the forthcoming discussion. The 
thermodynamic temperature $T_\text{th}$ is equal to $1/\beta$ when working in the canonical framework, or   
specified by the fundamental relation $(\partial s/ \partial e)^{-1}$, if the microcanonical viewpoint is instead adopted. 
A system and a thermal reservoir are said to be in equilibrium (no heat is exchanged) if their respective {\it thermodynamic temperatures} match. The canonical ensemble, as defined by the above LFT, provides hence the appropriate descriptive scenario for the system dynamics, $\beta$ being fixed as the inverse of the thermostat thermodynamic temperature.
If a difference in  temperature manifests between two adjacent systems, the heat flows against the temperature gradient so to eventually restore the equilibrium condition. This is the celebrated Fourier's law. With reference to both microcanonical and canonical pictures, one can alternatively consider the definition of kinetic temperature \cite{Balescu}, 
namely $T_\text{k}=\int p^2 \bar{f}(\theta,p) d \theta d p$.
For a Boltzmannian system $\bar{f}(\theta,p) \propto \exp(-\beta p^2/2)$ \cite{Pathria} and it is therefore immediate to see that $T_\text{k}=1/\beta= T_\text{th}$. When instead the system displays non-Boltzmannian traits, $T_\text{k} \ne T_\text{th}$.
Unintuitive phenomena can thus set in as the emergence of negative kinetic heat in the canonical ensemble, a possibility 
demonstrated in \cite{Staniscia}. Moreover, for a non-Boltzmannian fluid, a word of caution should be exercised on the usage of Fourier's 
heat law: does the heat flow across the kinetic or thermodynamic temperature gradient? We shall return later on this crucial point.  

\begin{figure}
\centering
\subfigure
{\resizebox{0.47\textwidth}{!}{ \includegraphics{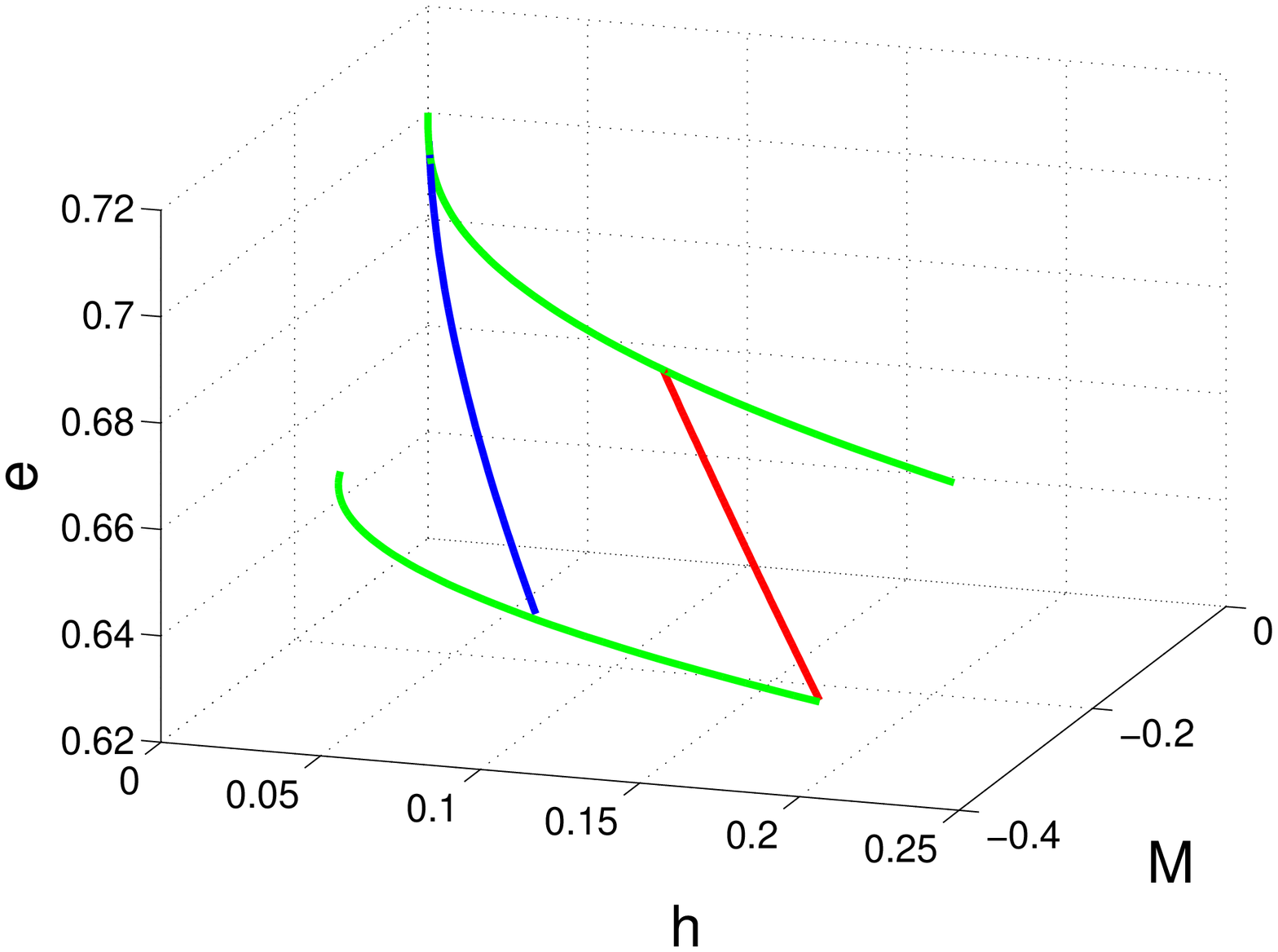}}}
\subfigure
{\resizebox{0.47\textwidth}{!}{ \includegraphics{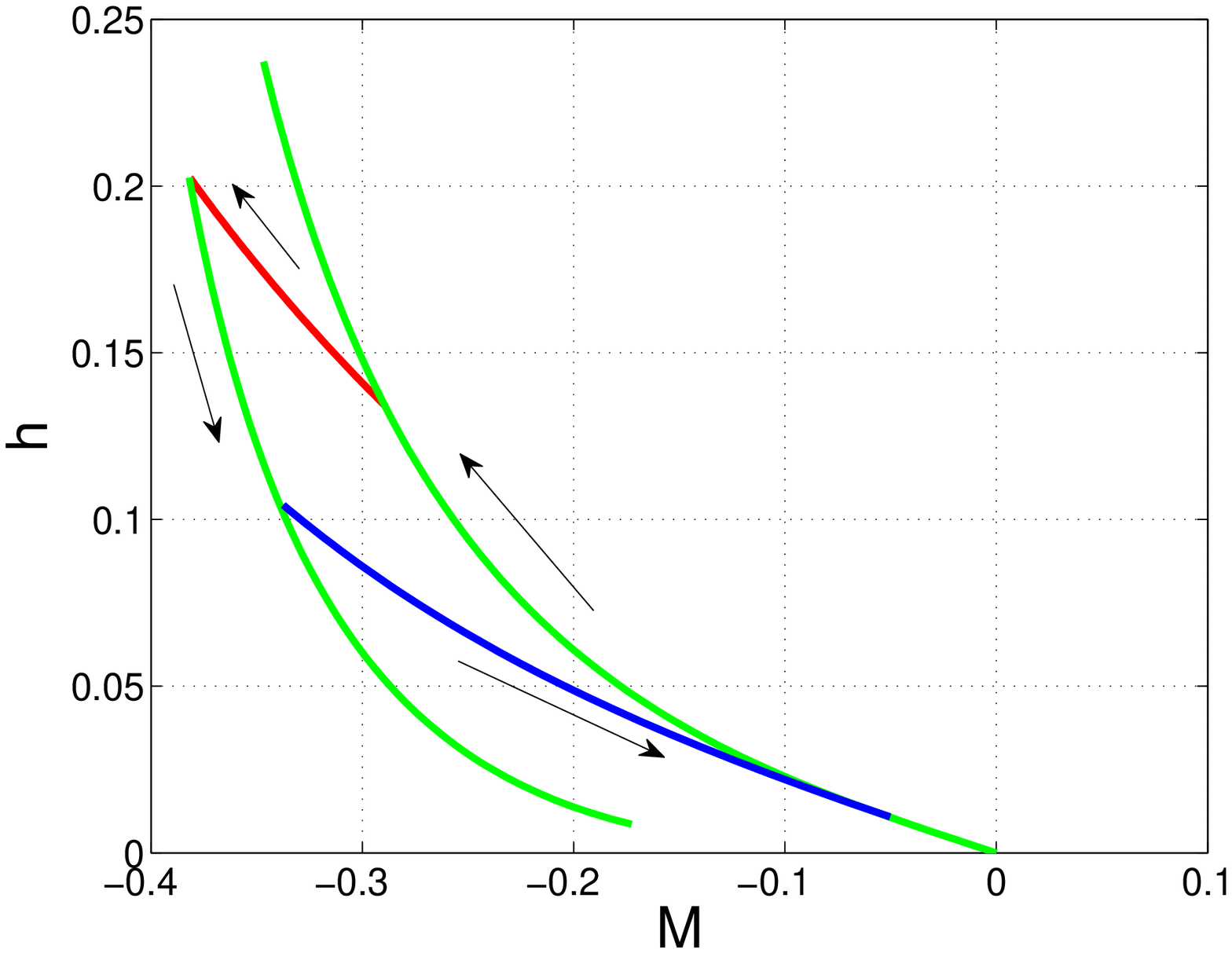}}}
   \caption{Upper panel: The thermodynamics cycle is represented in the three dimensional space $(h,M,e)$. The red line stands for the isothermal transformation carried out at $T_H$, while the blue line refers to the choice $T_C<T_H$. Here $T_H=1/\beta_H=0.33$ and $T_C=1/\beta_C=0.26$. The adiabatic transformations that close up the cycle are depicted in green. The heat exchanged with the reservoirs are respectively: $Q_H=-0.0363$ (heat flows out the system) and $Q_C=0.0416$ (the heat flows towards the system).  Here, $f_0=0.12$. 
Lower panel: The cycle is projected in the ($M,h$) subspace. The arrows indicate the direction of circulation that produces a positive work.} 
\label{fig1}
\end{figure}

{\bf Constructing the thermodynamic cycle.}
Having reviewed the definition of canonical and microcanonical ensembles, we can formally construct a thermodynamic cycle, reminiscent of Carnot's one, for an hypothetic device working with the Vlasov fluid introduced above. We will in particular consider two isothermal transformations, the reservoirs being characterized by the (thermodynamic) temperatures $T_H$ (hot) and $T_C$ (cold). The cycle is then closed by assuming two adiabatic transformations. As concerns the latter transformation, we recall \cite{Amit} that the first principle of thermodynamics reads:
\begin{equation}
\delta Q = d e - h d M.
\label{amit}
\end{equation}
Here, $Q$ is the heat provided by the environment to the system. The adiabatic transformation $\delta Q=0$ is hence realized by tuning the field strength as $h=d e/ d M$ along the path. A representative cycle obtained by implementing the procedure outlined above is depicted in figure \ref{fig1}. Each point displayed in the three dimensional space ($h,M,e$) is either a stationary solution of the entropy functional $s$ (microcanical, adiabatic lines) or of the free energy $\phi$ (canonical, isothermal). Notice that one needs to circulate the cycle in figure \ref{fig1} (lower panel) counter-clockwise, so to extract a positive work amount  ($L_{HMF}=-\int_{cycle} h dM$). The heat exchanged with the hot and cold reservoirs, labeled respectively $Q_H$ and $Q_C$, can be readily evaluated as an application of the first principle (\ref{amit}) and upon calculation of the work performed along the isothermal tracks. Surprisingly, and as a straightforward application of the above procedure, we obtain $Q_H <0$ and $Q_C>0$. Moreover, $Q_C > |Q_H|$. This is at odds with the physical constraints imposed by the second law of thermodynamics. The virtual machine that we have constructed works in fact as a refrigerator, cooling down the reservoir with the lower temperature. In doing so, it however generates a {\it positive} work load, which violates Clausius formulation of the second law of thermodynamics. In principle, and admitting this conclusion to be correct, one could insert between the two considered reservoirs a standard thermal machine that, in a cycle, receives from the hotter bath the heat quota $Q_H$ (i.e. the very same released by our Vlasov device), and returns to the colder reservoir the heat $Q_C'$, such that $|Q_C'|<Q_H$, while producing a positive work $L$.  The machine obtained by pairing together the two considered devices, subtracts heat from a single reservoir ($Q_C-|Q_C'|>0$) and returns the positive work $L_{tot} = L_{HMF}+L$, against the second law of thermodynamics in the Kelvin-Planck formulation. In other words, the machine that we have assembled, and which intimately exploits the unintuitive features of long range systems, is characterized by an efficiency $\eta=L_{tot}/(Q_C-|Q_C'|)$ equal to one. This is so unexpected that, at first sight, it would be tempting to reject the results as a whole. It is however possible to elaborate on a thermodynamic interpretation of the results and pinpoint to the specific, and highly unconventional phenomenon, that is ultimately responsible for such an astonishing behavior. Then we will come back to discussing the foundation of the theory, especially the customarily adopted definition of canonical ensemble, and suggest that the validity of the second principle could be eventually restored provided 
the kinetic temperature and {\it not} the thermodynamic one is employed in the law of Fourier for the heat transport.  

{\bf Negative kinetic heat capacity and the violation of the second principle of thermodyamics.}

The cycle reproduced in figure \ref{fig1} extends within a domain of the parameters space yielding magnetized states. In \cite{Staniscia}, it was shown that negative (kinetic) specific heat is found in the canonical ensemble, i.e. along the isothermal lines of the cycle, when the magnetization is different from zero. This fact is indeed responsible for the apparently unphysical conclusions reached above. Consider in fact the first law and isolate the two terms that enter the right hand side of equation (\ref{amit}). The second term is related to the actual physical work: when proceeding counter-clockwise along the isothermal transformation, in contact with the hot reservoir, the quantity $-h dM$ is positive as can be deduced by looking at figure \ref{fig1} (lower panel, red line). In  practice, and as it happens for an ideal gas performing an isothermal transformation, the work applied perturbes the system from its equilibrium condition, while constraining it to the same thermodynamic temperature of the bath.  A (positive) heat quota ($-hdM$) is consequently transferred from the bath to the system, when circulating counter-clockwise. The opposite clearly holds when the the cold isothermal is considered, together with the chosen versus of circulation: $-hdM$ is negative, see figure \ref{fig1} (upper panel, blue line). While the thermodynamic temperature is kept constant, the kinetic one $T_k$ changes along the considered transformation: in particular, $T_k$ increases along the isothermal at  $T_H$, decreases when the system is evolved in contact with the colder bath $T_C$.  

The other contribution to the heat $\delta Q$, the first term ($d e$) on the right hand side of equation  (\ref{amit}), gets also modulated along the transformation. In figure \ref{fig2}, the energy $e$ is shown to decrease when $T_k$ increases, for both cold and hot isothermals (resp. red and blue). The kinetic specific heat is thus negative, a condition that realises because of the non-Boltzmannian character of the system. Increasing the kinetic temperature, as the system does when in contact with the hot bath $T_H$, implies decreasing the energy $e$, and so contributing with a negative term to the global heat exchanged with the reservoir. Incidentally, the latter contribution turns out to be larger in absolute value, than the positive contribution associated to the magnetic work. The total contribution is hence negative and for this reason the system cedes heat to the hot reservoir. Similar conclusions can be drawn when looking at the other isothermal transformation: now $d e$ is positive (the kinetic temperature decreases) and overcomes the negative $-hdM$ contribution. 
When kinetic and thermodynamics temperature are equal (as they should for conventional Boltzmannian short range systems) the (kinetic) specific heat has to be necessarily positive, and the above surprising condition cannot realize.

{\bf Discussion.}

Summing up we have analytically shown that non-Boltzmannian systems can apparently violate the second law of thermodynamics, and explained such an astonishing ability in terms their peculiar statistical mechanics, i.e. existence of negative kinetic specific heat in the canonical ensemble. Importantly, and as a direct consequence of the above analysis, a cyclic thermal device working with a non-Boltzmannian fluid can be imagined which has efficiency equal to one. It is however hard to accept such a conclusion, which would revolutionize our current understanding of the laws of nature. 

\begin{figure}
\centering
{\resizebox{0.47\textwidth}{!}{ \includegraphics{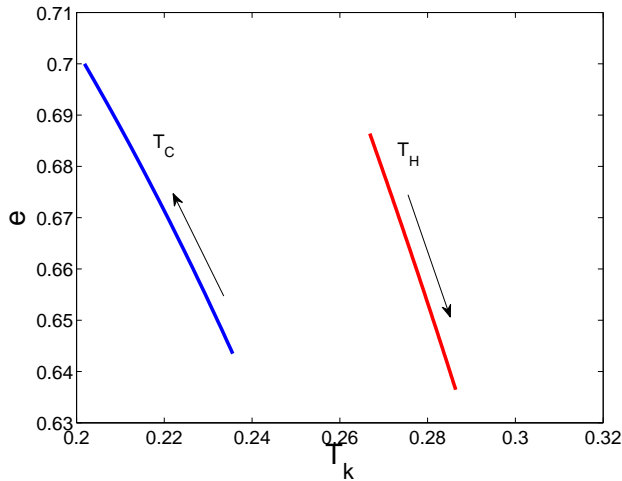}}} 
\caption{The energy  $e$ of the system is plotted versus the kinetic energy $T_k$, along the two isothermal transformations (in red $T_H=0.33$, in blue 
$T_C=0.26$). The transformations proceed along the directions highlighted by the arrows. Here, $f_0=0.12$. 
\label{fig2}}
\label{fig2}
\end{figure}
A way out in our view exists which potentially allows one to respect the second principle and confirm its validity for the class of systems being here considered. The critical point has to do with the definition of the canonical ensemble, that we have here mutuated from equilibrium classical statistical mechanics, by performing the Legendre-Fenchel Transform of the entropy functional. In doing so, one requires that the canonical dynamics preserves the thermodynamic temperature of the system setting it to a constant value imposed by the external reservoir. Alternatively, one could imagine that 
non-Boltzmannian systems, when placed in contact with a thermal bath, would organize so to freeze their {\it kinetic temperature} to the temperature \footnote{Kinetic temperature, in general. If one deals with a Boltzmannian bath,  kinetic and thermodynamics temperature clearly coincide.} of the bath. As a natural consequence, already stressed before, 
the Fourier law \cite{Lebowitz} should  require that the heat flows along the {\it kinetic temperature gradient}, and not along the thermodynamic one, as the definition of canonical thermodynamical equilibrium would indirectly imply. Pushing further this proposal, we could possibly ends up with a cyclic non-Boltzmannian engine which    
meets the second law requirement, while still exceeding the Carnot upper bound of efficiency, since the thermodynamic temperature is related to the kinetic one by  a non linear functional relation \cite{Staniscia}. This is an important issue that touches the foundation of modern physics and that deserves to be carefully addressed. 

To conclude, we would like to stress that our study is relevant for a quite large class of systems and for many practical physical situations. In fact, the Vlasov equation is often invoked as the reference model for several  
experimental systems, involving a  large number of individual  components, e.g. charged systems, plasmas \cite{Dubin}, free-electron lasers \cite{fel} and magnetic dipolar systems \cite{dip}.
In particular, it can be shown that, under some hypotheses, the specific system here considered can be formally 
reduced, in the limit for $h=0$, to that governing the dynamics of collective atomic recoil lasers (CARL) \cite{carl, Bachelard}. 
This important observation could open up the perspective for an experimental verification of our theoretical predictions.

\end{document}